\documentclass[preprint,12pt]{elsarticle}

\usepackage{graphicx}
\usepackage{pdfpages}
\usepackage{subfigure}
\usepackage{float}
\usepackage{lineno}
\usepackage{amsmath}
\usepackage{multirow}

\journal{Nuclear Instruments and Methods in Physics Research A}

\biboptions{sort&compress}

\begin{document}
	
\begin{frontmatter}
		


\title{Development of low-radon ultra-pure water for the Jiangmen Underground Neutrino Observatory}

\renewcommand{\thefootnote}{\fnsymbol{footnote}}
\author{
	T.Y.~Guan$^{a}$,
	Y.P.~Zhang$^{b,c,d}$,
	B.~Wang$^{a}$,
	C.~Guo$^{b,c,d}\footnote{Corresponding author. Tel:~+86-01088236256. E-mail address: guocong@ihep.ac.cn (C.~Guo).}$,
	J.C.~Liu$^{b,c,d}$,
	Q.~Tang$^{a}\footnote{Corresponding author. Tel:~+86-13974753537. E-mail address: tangquan528@sina.com (Q.~Tang).}$,
	C.G.~Yang$^{b,c,d}$,
	C.~Li$^{a}$
}
\address{
	${^a}$ School of Nuclear Science and Technology, University of South China, Hengyang, 421001, China\\
	${^b}$ Experimental Physics Division, Institute of High Energy Physics, Chinese Academy of Sciences, Beijing, 100049, China \\
	$^{c}$ School of Physics, University of Chinese Academy of Sciences, Beijing, 100049, China \\
	$^{d}$ State Key Laboratory of Particle Detection and Electronics, Beijing, 100049, China
}


		
\begin{abstract}
The Jiangmen Underground Neutrino Observatory(JUNO) is a state-of-the-art liquid scintillator-based neutrino physics experiment under construction in South China. To reduce the background from external radioactivities, a water Cherenkov detector composed of 35~kton ultra-pure water and 2,400 20-inch photomultiplier tubes is developed. Even after specialized treatment, ultra-pure water still contains trace levels of radioactive elements that can contribute to the detector background. Among which $^{222}$Rn is particularly significant. To address this, an online radon removal system based on the JUNO prototype has been developed. By integrating micro-bubble generators to enhance degasser's radon removal efficiency, the radon concentration in water can be reduced to 1~mBq/m$^{3}$ level, meeting the stringent requirements of JUNO. Additionally, a highly sensitive online radon concentration measurement system capable of detecting concentrations $\sim$1~mBq/m$^3$ has been developed to monitor the radon concentration in water. In this paper, the details regarding both systems will be presented.

\end{abstract}


\begin{keyword}
	
	
Ultra-pure water\sep Radon\sep Degassing membranes\sep Microbubble generator	

\end{keyword}

\end{frontmatter}

\section{Introduction}

The Jiangmen Underground Neutrino Experiment (JUNO)~\cite{adam_juno_2015,noauthor_juno_2022} is a significant endeavor in the field of neutrino physics, representing the second neutrino experiment led by China after the successful implementation of the Daya Bay experiment~\cite{dayabay}. JUNO is a 20~kton Liquid Scintillator (LS) detector under construction in a laboratory located 700~m underground in Jiangmen City, Guangdong Province. An excellent energy resolution and a large fiducial volume offer exciting opportunities for addressing many important topics in neutrino and astroparticle physics. With six years of data taking, JUNO will determine the neutrino mass ordering at 3-4~$\sigma$ significance as well as the neutrino oscillation parameters sin$^2$$\theta_{12}$, $\Delta$m$^2_{21}$, and $\mid \Delta$m$^2_{32}\mid$ to the precision of 0.6\% or better by detecting reactor antineutrinos from the Taishan and Yangjiang nuclear power plants. Moreover, JUNO can also observe neutrinos from other sources, including supernova burst neutrinos, diffuse supernova neutrino background, geo-neutrinos, atmospheric neutrinos, and solar neutrinos~\cite{adam_juno_2015}.

\begin{figure}[htb]
\centering
\includegraphics[height=6cm]{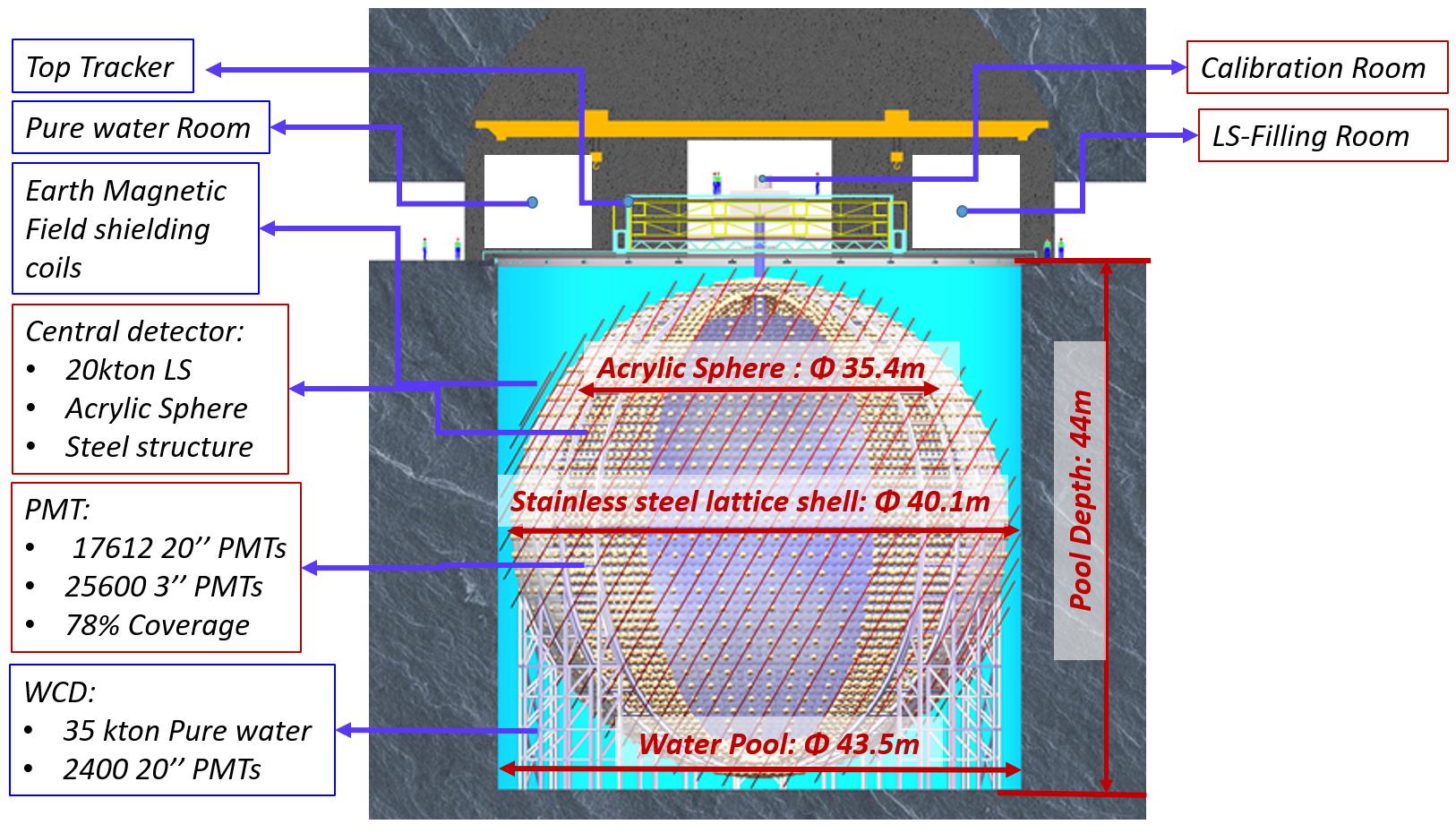}
\caption{Scheme of the JUNO detector. }
\label{fig.JUNO}
\end{figure}

Fig.~\ref{fig.JUNO} shows the scheme of the JUNO detector. The target, 20~kton of ultra-pure LS, is contained in a spherical acrylic vessel. The light emitted by LS is watched by 17612 20-inch photomultiplier tubes (PMTs) and 25600 3-inch PMTs. To shield the radioactive background from the surrounding rocks, the entire LS detector is submerged in a Water Cherenkov Detector (WCD). The WCD is a cylinder of 43.5~m in diameter and 44~m in height, which is filled with 35~ktons of ultrapure water. Although ultra-pure water has been treated with specialized equipment, it still contains trace levels of radioactive elements. Radon is one of the most important radioactive background sources. To lower the accidental background in the central detector, the radon concentration in the ultra-pure water should be reduced to less than 10~mBq/m$^3$~\cite{adam_juno_2015}. To remove radon from water and to monitor the radon concentration in water, a radon removal system and a radon concentration measurement system based on the water system of the JUNO prototype have been developed. 

This paper is organized as follows. Sec.2 describes the ultra-pure water system of the JUNO prototype, sec.3 describes the online radon concentration in water measurement system, sec.4 describes the measurement results and sec.5 is the summary.

\section{Ultra-pure water system of the JUNO prototype}

\begin{figure}[htb]
	\centering
	\includegraphics[height=5.5cm]{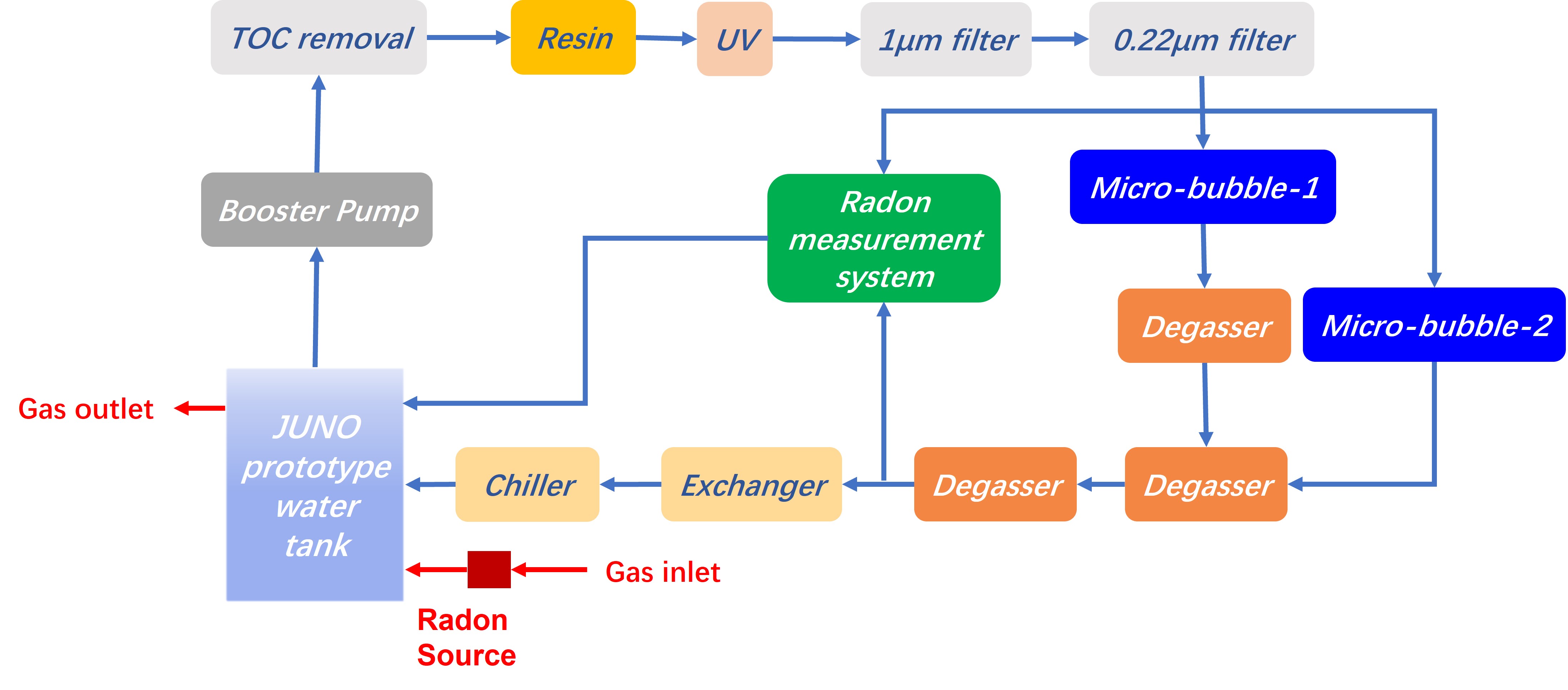}
	\caption{Flow Chart of the water system of the JUNO prototype.}
	\label{fig:watersystem}
\end{figure}

Fig.~\ref{fig:watersystem} shows the flow chart of the water system of the JUNO prototype. The booster pump is employed for powering water circulation, Total Organic Carbon (TOC) removal is utilized to eliminate organic matter from water, the resin (DOWEX$^{TM}$ MONOSPHERE$^{TM}$ MR-3 UPW grade~\cite{resin}) is employed for removing ions from water, the UV, with a wavelength of $\sim$254~nm, is utilized for water sterilization, the filters are employed for particulate matter removal, the microbubbles are utilized for gas-loading, the degassers are utilized for radon removal, the exchanger is used for heat exchange, and the chiller is to regulate the water temperature to its designated value of 20 $\pm$ 1~$^{\circ}$C. While evaluating the radon removal efficiency, a radon source is utilized to introduce radon gas into the water to maintain a stable radon concentration within the water, and evaporating nitrogen is used as the gas source. After passing through the filter, a small portion of the water is separated and directly introduced to the second stage degasser. The primary objective is to increase the gas concentration in the water entering the second stage degasser, thereby enhancing its radon removal efficiency.

The degassers are  Liqui-Cel degassing membranes~\cite{Li-cel}, which are widely employed technology in the field of water treatment and fluid processing. Two stages of X-50 membranes and one stage of  X-40 membrane are installed and serial-connected in the water circulation system. The micro-bubble generators are manufactured by RuJing Environmental Protection Company. The maximum water flow rate for Micro-bubble-1 is 1000~L/h, whereas that of Micro-bubble-2 is 300~L/h.

\begin{figure}[htb]
	\centering
	\includegraphics[height=5.5cm]{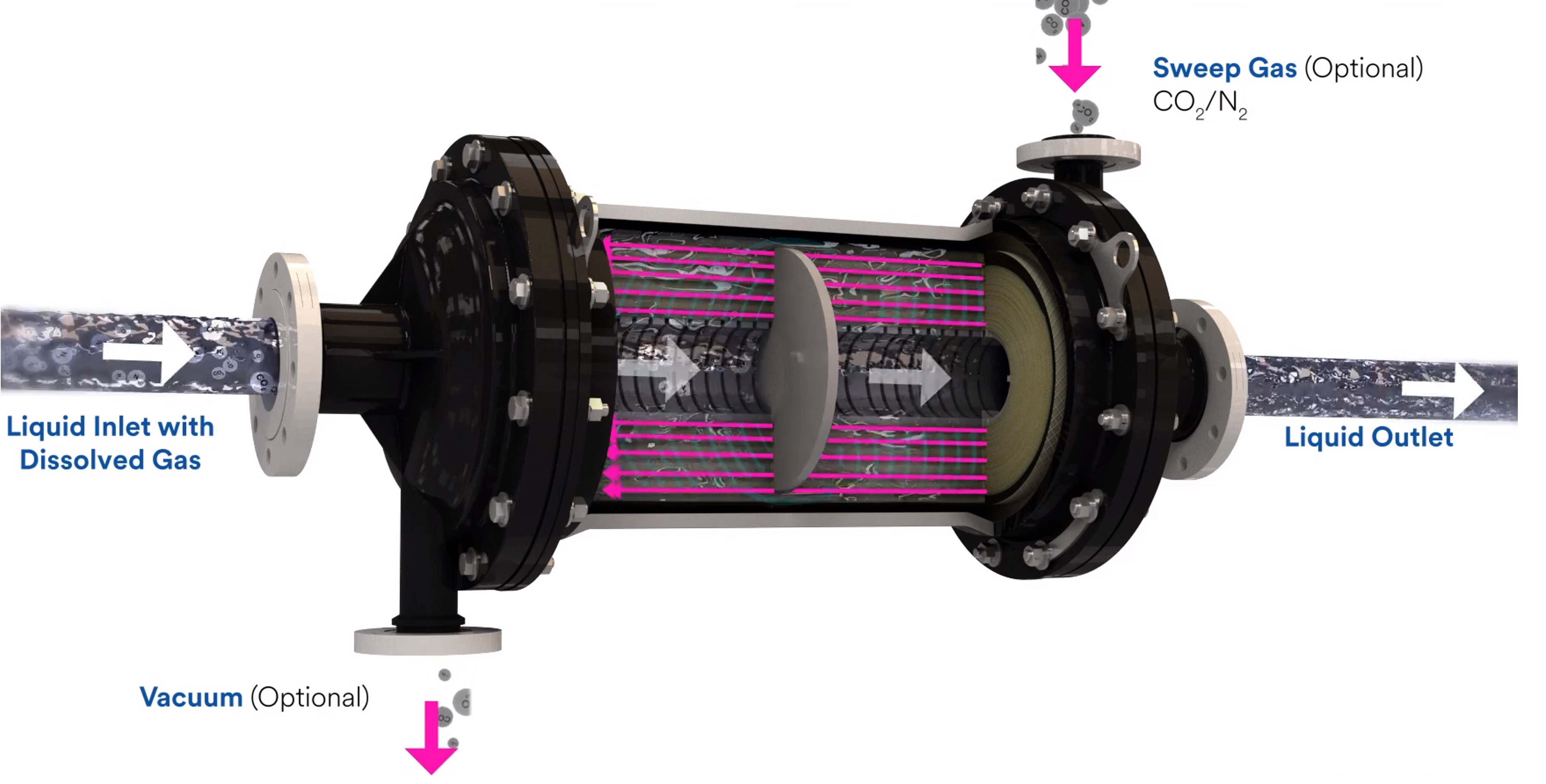}
	\caption{Schematic diagram of the degasser. The vacuum and sweep gas options are flexible, allowing for the use of either one specifically or both simultaneously. After testing and comparing, we employed the vacuum mode in conjunction with 3~L/min of evaporated nitrogen as the sweep gas in this experiment.}
	\label{fig:degasser}
\end{figure}

Fig.~\ref{fig:degasser} shows the diagram of the degasser~\cite{Li-cel}, which is comprised of an intricate arrangement of microporous polypropylene hollow fibers wound around a central tube. These hollow fibers are evenly spaced to optimize the flow capacity and the utilization of the total membrane surface area. The hydrophobic nature of the hollow fiber membrane prevents liquid penetration into the membrane pores. When high-pressure liquid flows over the outer shell side of the hollow fibers, a vacuum is applied to the inner side to create the driving force for dissolved gas in the liquid to pass through the membrane pores. The degassing membrane device exhibits a remarkable efficiency of up to 90\% in removing oxygen and carbon dioxide~\cite{Li-cel}, thereby demonstrating its potential for gas removal. Considering the solubility of radon gas in water, this device also showcases a notable capability in efficiently eliminating radon gas~\cite{membrane_2017}. According to our previous work~\cite{guo_study_2018}, the degasser's radon removal efficiency is related to the gas concentration in the water, when the gas concentration in water decreases, the degasser's radon removal efficiency decreases. An effective way to increase the degasser's radon removal efficiency is to increase the gas concentration in the water, and thus microbubble generators are introduced into the water system.

A microbubble generator is a device that produces bubbles on a micrometer scale through vibration or pressurization, leading to the dissociation of gas molecules from the gas-liquid interface and the formation of bubbles in the liquid phase. Microbubbles, characterized by their small size, comparatively large surface area, high adhesive efficiency, and slow ascent speed in water, play a vital role in various application fields such as air flotation, water purification, water reoxygenation, pharmaceuticals, and precise chemical reactions~\cite{MB-1,MB-2}. In this study, the microbubble generator is utilized to enhance gas concentration in the water~\cite{2012Microbubble}. With the same volume of air, the microbubble count is significantly higher, resulting in an increased total contact area between the bubbles and the water due to the corresponding increase in bubble surface area. Considering this augmented surface area, the dissolution capacity of microbubbles is approximately 200,000 times higher than that of regular bubble~\cite{2006Ozone, liu_degradation_2018,khan_micronanobubble_2020}. 

\begin{figure}[htb]
	\centering
	\includegraphics[height=5.5cm]{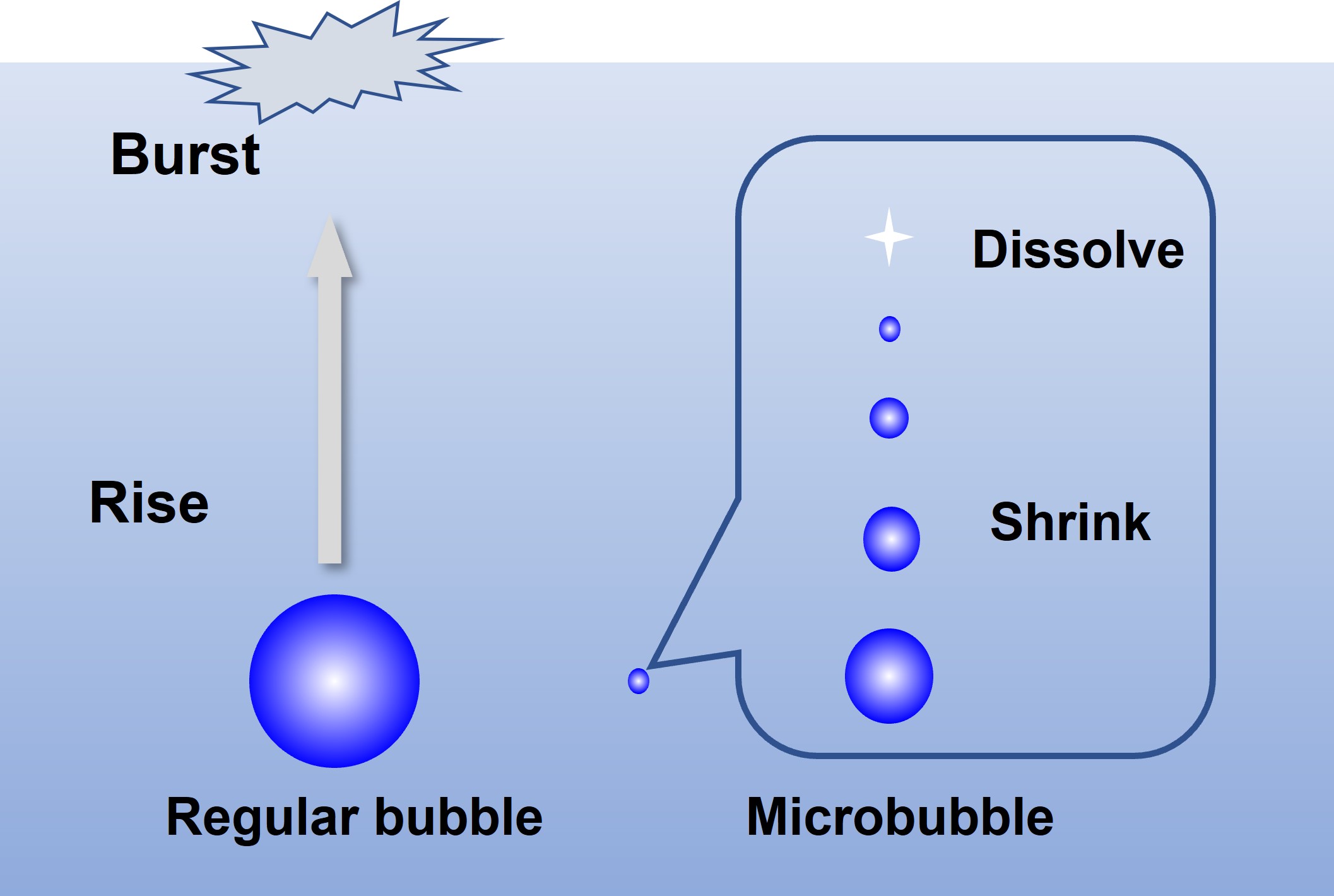}
	\caption{Behavior comparison of regular bubbles and microbubbles in water.}
	\label{fig:microbubble}
\end{figure}

\begin{figure}[htb]
	\centering  
	\subfigure[]{\includegraphics[width=0.5\linewidth,angle=270]{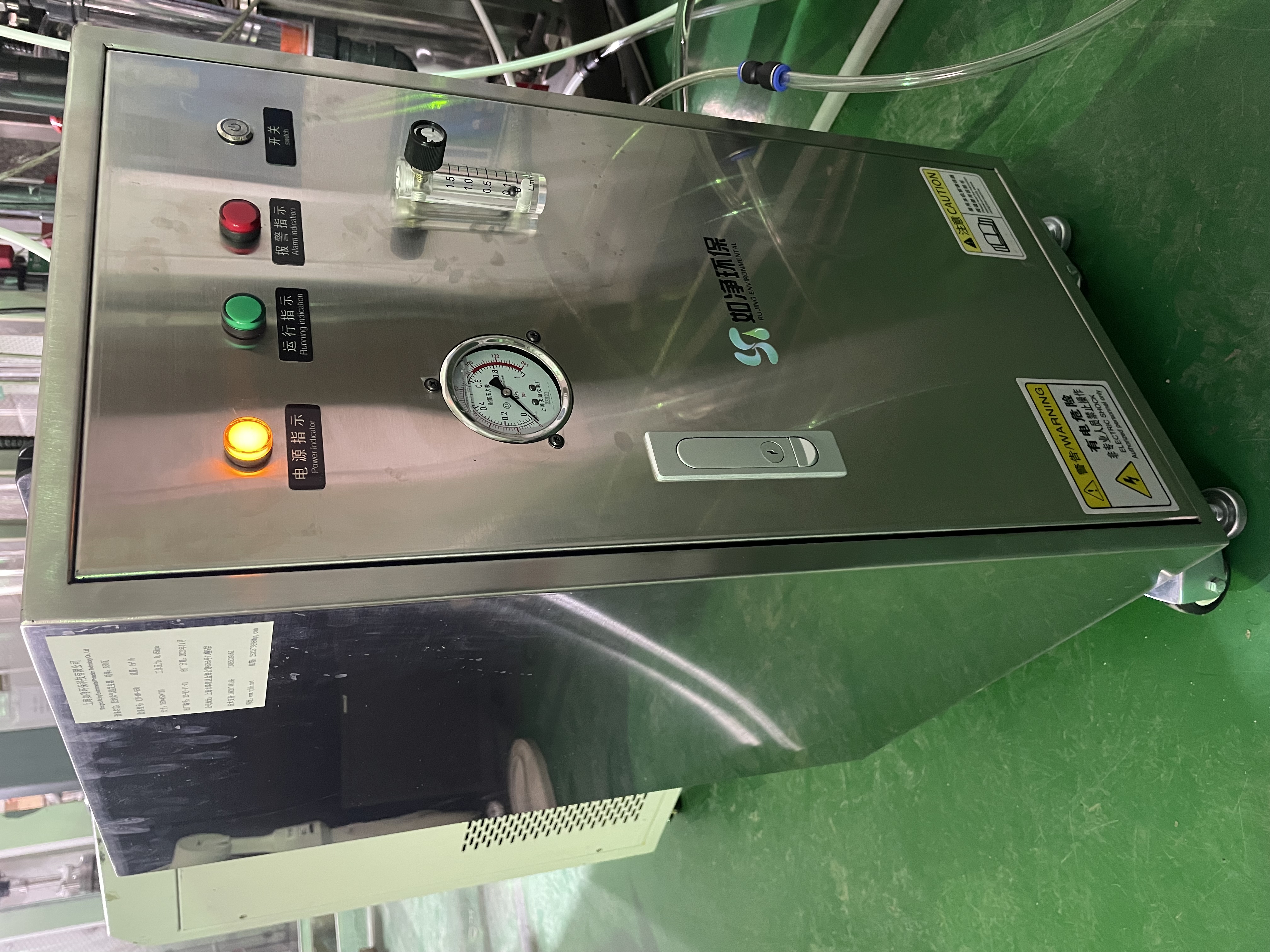}}
	\subfigure[]{\includegraphics[width=0.5\linewidth,angle=270]{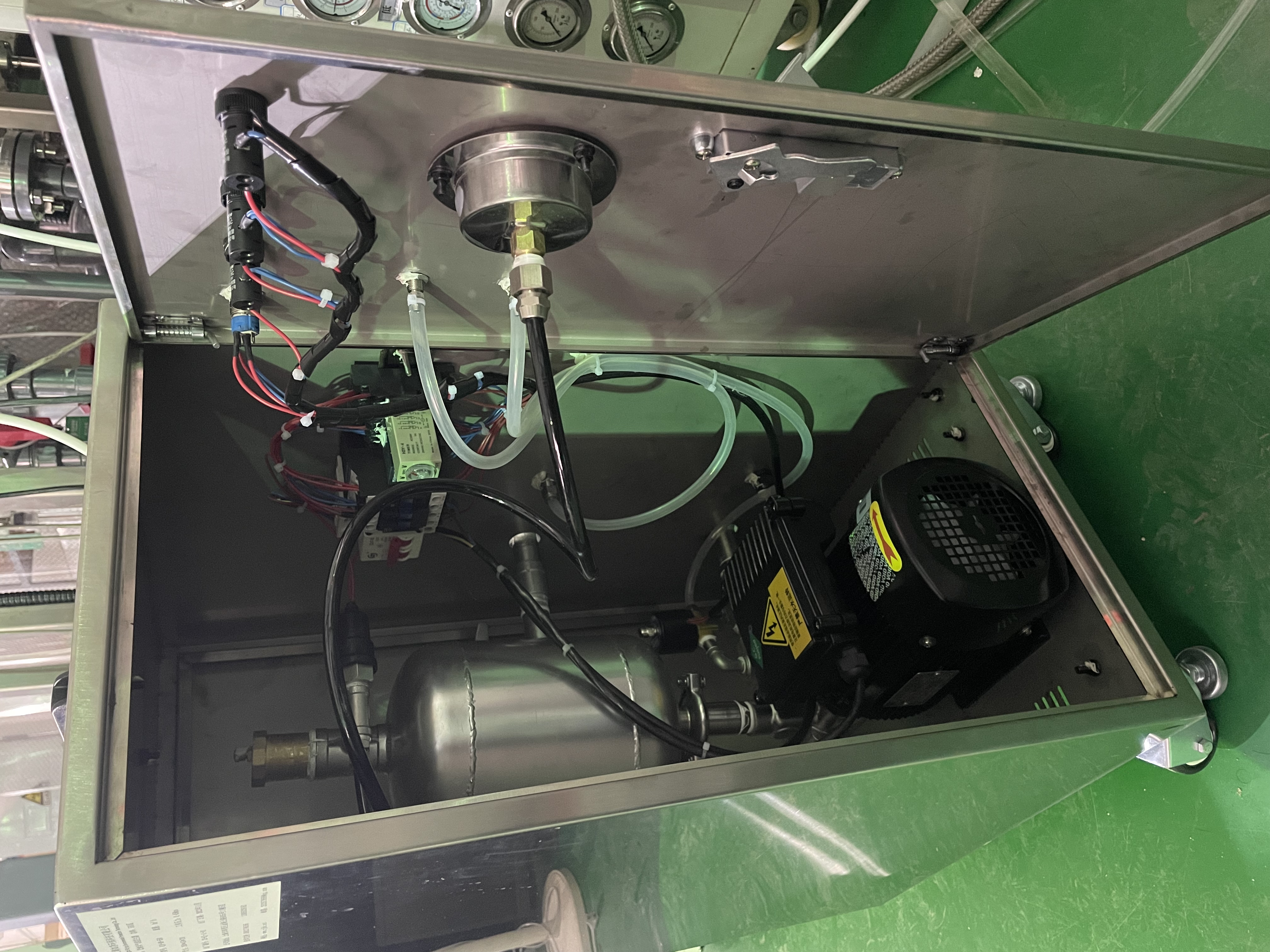}}
	\caption {Pictures of the microbubble generator used in this study. }
	\label{Micro_bubble}
\end{figure}

Fig.~\ref{fig:microbubble} compares the behavior of regular bubbles and microbubbles in water. While regular bubbles rise in water and subsequently burst on the surface, microbubbles shrink and gradually dissolve in water. Fig.~\ref{Micro_bubble} shows pictures of the microbubble generator used in this study. The gas source for the microbubble generators is evaporating nitrogen, in which the radon concentration is $\sim$0.1mBq/m$^3$~\cite{liu_system_2023}. The ratio of gas to water flow rates is optimized in this study, which will be discussed in Sec.4.3.

\section{Online radon concentration in water measurement system}
\subsection{System composition}
\begin{figure}[htb]
	\centering
	\includegraphics[height=6cm]{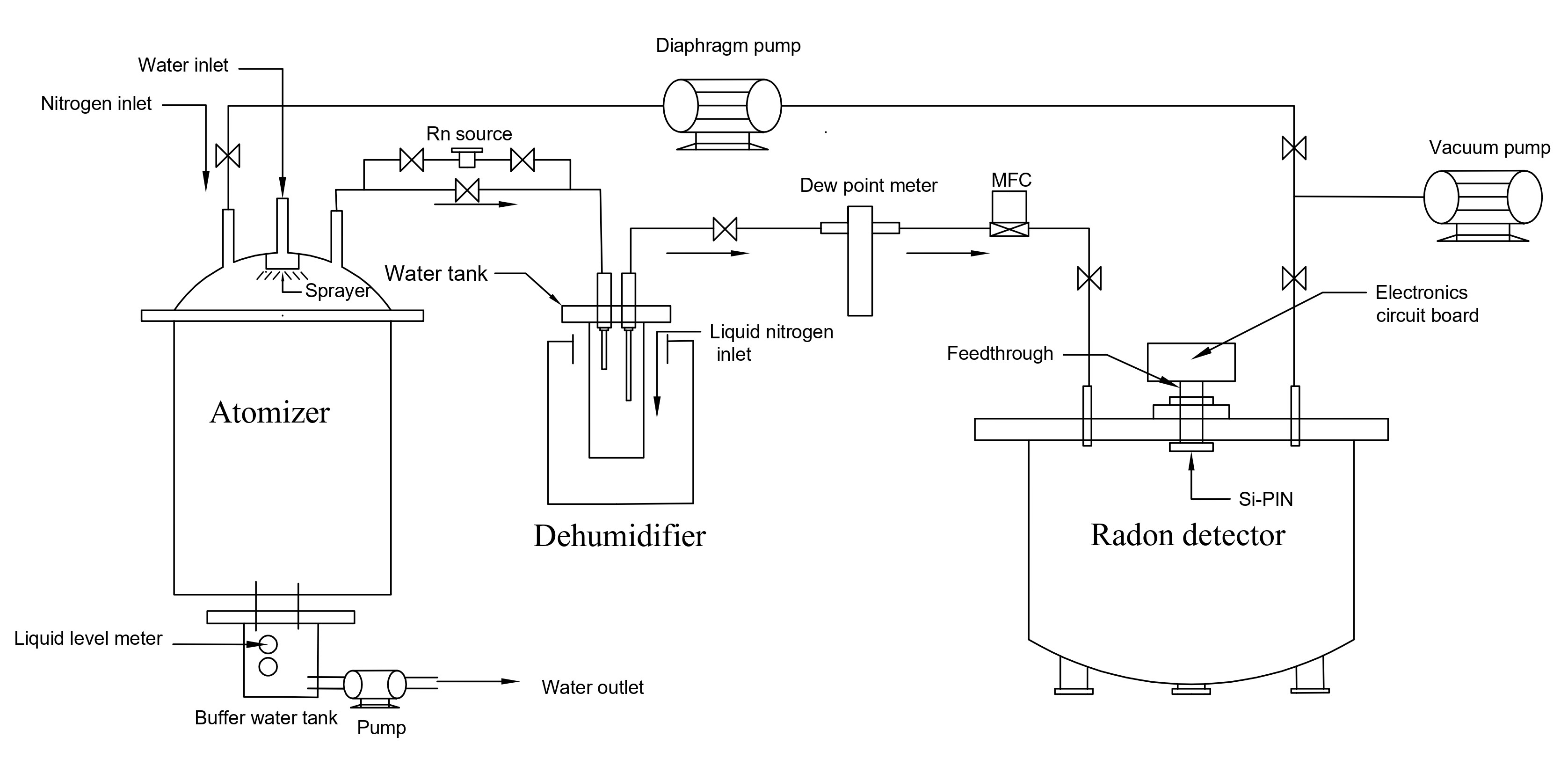}
	\caption{Schematic diagram of the radon concentration in water measurement system.}
	\label{fig:radonsystem}
\end{figure}

Fig.~\ref{fig:radonsystem} shows the schematic diagram of the radon concentration in water measurement system, which consists of an atomizer, a dehumidifier, a radon detector, and a set of gas circulation system. This system operates in two modes: recirculation mode and gas transfer mode. In the recirculation mode, gas continually circulates between the atomizer and the detector, requiring the diaphragm pump and dehumidification system to remain operational at all times. This mode offers the advantage of real-time monitoring of radon concentration in water. However, it has the disadvantage of requiring a large amount of liquid nitrogen for the dehumidification system, and frequent opening of the dehumidification system's tank to remove the condensed water. In the transfer mode, water-vapor equilibrium is first achieved in the atomizer, followed by evacuation of the radon detector. The pressure difference is then utilized to transfer the gas from the atomizer to the detector. This mode offers the advantage of requiring a smaller amount of liquid nitrogen and condensed water. However, the drawback is that real-time monitoring of radon concentration in the water is not possible. The details and functions of each device are as below.

A) Atomizer. 

The radon concentration in water can not be measured directly, it has to be transferred into gas. The atomizer with a volume of $\sim$95~L serves as a water vapor equilibrium device, facilitating the conversion of radon from water into vapor through spraying. The sprayer, located at the top of the atomizer, is connected to the sample port of the water system. A buffer water tank is situated at the bottom of the atomizer to collect the water sprayed down by the sprayer. Additionally, an inductive high and low level meter is attached to the exterior of the buffer water tank, which controls the pump to open and transfer water into the JUNO prototype water tank when the water level exceeds the upper side of the water level meter. When they reach equilibrium, the radon concentration ratio in water to vapor R can be calculated according to the temperature-dependent air/water equilibrium Fritz-Weigel equation~\cite{Weigel,RM,JER}: 
\begin{equation}
R={0.105+0.405e^{-0.0502T}}
\label{Eq.R}
\end{equation}
where T is the temperature in the unit of centigrade. In the case of the JUNO prototype, the water temperature has been maintained at $\sim$20~$^{\circ}$C~\cite{adam_juno_2015}, resulting in an R value of $\sim$0.25. Consequently, the radon concentration in water can be determined based on the gas measurement results.

B) Radon detector. 

A highly sensitive radon detector is utilized to measure the radon concentration in the gas. The radon detector comprises a 100~L cylindrical electro-polished stainless steel vessel and a Si-PIN photodiode. The principle of the radon detector is based on electrostatic collection. A negative high voltage is applied to the surface of the Si-PIN, while the stainless steel chamber is grounded, generating an electric field from the stainless steel chamber to the Si-PIN. As over 90\% of $^{222}$Rn daughters are positively charged~\cite{SuperK1999,10.1093/ptep/ptv018,SNO}, they can be collected onto the surface of the Si-PIN. Then the $^{222}$Rn concentration can be determined by detecting the $\alpha$s from the decay of $^{214}$Po and $^{218}$Po.

\begin{figure}[htb]
	\centering  
	\includegraphics[height=4.5cm]{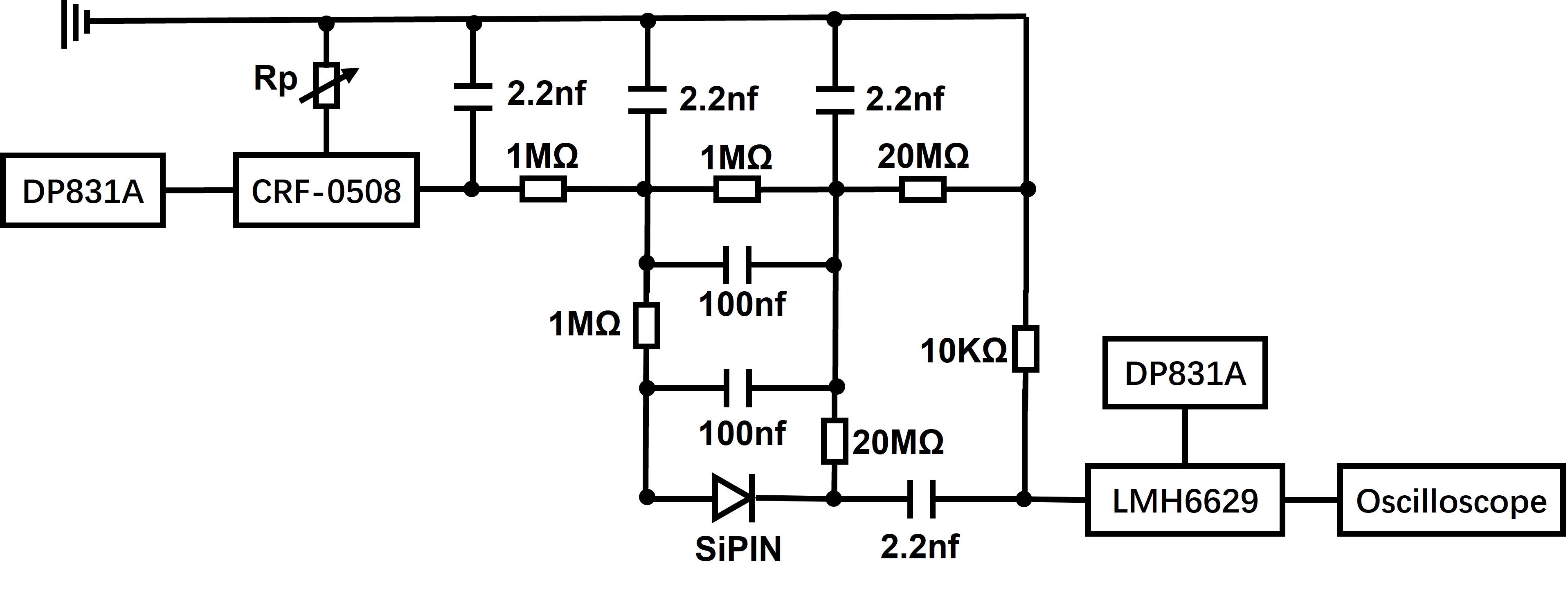}
	\caption{The circuit diagram of the readout system. }
	\label{electronics}
\end{figure}

\begin{figure}[htb]
	\centering  
	\subfigure[]{\includegraphics[width=0.49\linewidth]{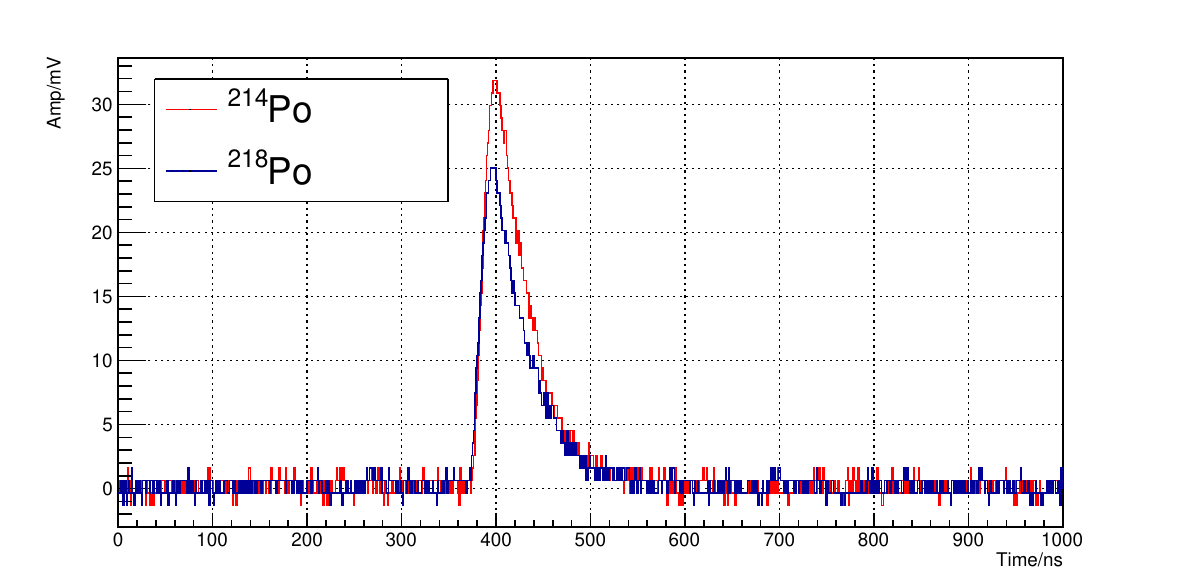}}
	\subfigure[]{\includegraphics[width=0.49\linewidth]{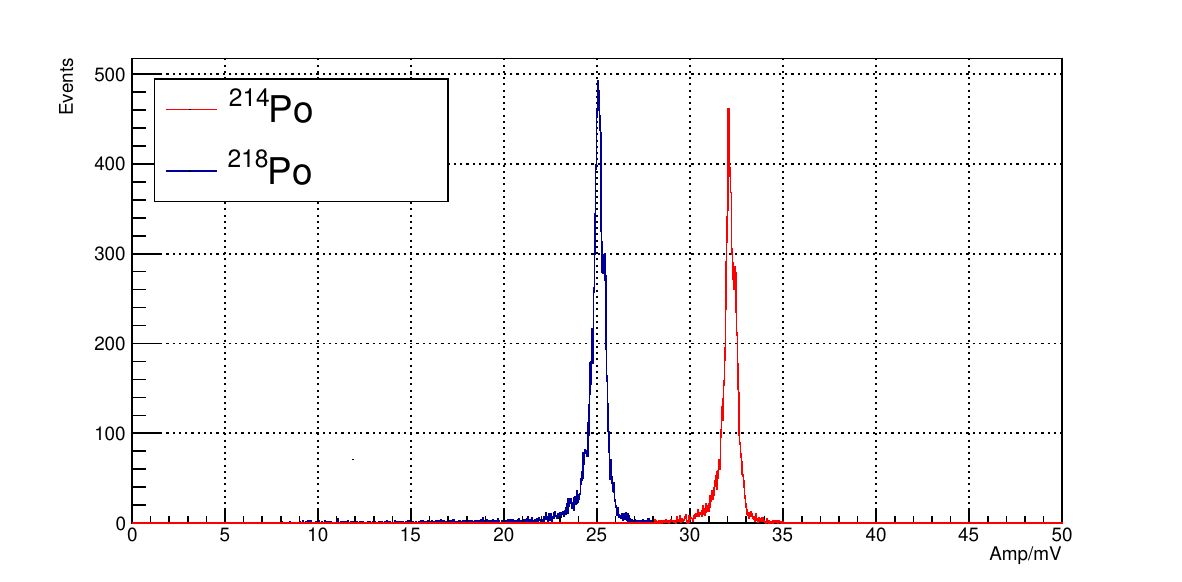}}
	\caption{Example of pulses of $^{218}$Po and $^{214}$Po signal and amplitude spectrum. (a) The pulses of  $^{218}$Po and $^{214}$Po signal. (b) A typical amplitude spectrum of $^{218}$Po and $^{214}$Po signal. }
	\label{signal}
\end{figure}

Compared with our previous works~\cite{zhang_development_2018,chenYY_2022,Li_2023,liu_system_2023}, the electronics have been updated. Fig.~\ref{electronics} shows the diagram of the readout system. For the high-voltage application, a DC-DC module (CRF-0508, Tianjin Sentell Company~\cite{HV_tianjin}) is employed. By adjusting the value of R$_p$, this module can convert a 12~V input voltage to an output voltage ranging from -700~V to -2200~V. For signal amplification, an amplifier board based on LMH6629 (TI company) is utilized. Two DC power suppliers (DP831A Rigol Technology Co.) are used to provide power to the DC-DC module and the amplifier. The primary advantage of this version over the previous electronics is the capability to remotely activate and deactivate the DC-DC module or the LMH6629 amplifier by controlling its power switch remotely. For the data acquisition, a desktop digitizer (DT5751, CAEN) is used to record the pulses. Example of pulses and the measured spectrum are shown in Fig.~\ref{signal}.

C) Dehumidifier.

The dehumidifier consists of a liquid nitrogen tank, a dewar, a water tank, a temperature controller, a temperature sensor (Pt100), and an electromagnetic valve. The Pt100 temperature sensor which is placed in the dewar serves as the input to the temperature controller. The temperature controller controls the switch of the electromagnetic valve, to realize the intermittent injection of liquid nitrogen into the dewar. The liquid nitrogen vaporizes in the dewar and the cooled nitrogen gas can cool down the measured gas. This system can realize a temperature change inside the dewar from room temperature to -196℃ and the temperature stability is within ±5~℃.  The details of the dehumidifier can be found in Ref.~\cite{wang_SiPM_2021}. A dew point meter is connected to the outlet of the dehumidifier to measure the humidity of the gas. One drawback of this system is the need for regular replenishment of liquid nitrogen. Consequently, an electrically cooled refrigerator will be used to replace the liquid nitrogen tank and Dewar shortly. For a balance of cost and performance, we have customized the refrigerator to maintain a constant temperature of -60~$^{\circ}$C. Therefore, in this study, the temperature of the dehumidifier is set to -60~$^{\circ}$C.

D) Gas circulation system.

The gas circulation system consists of a diaphragm pump (N022AT, KNF), a Mass Flow Controler (MFC, 1179A, MKS), a vacuum pump (ACP40, Pfeiffer), and the relevant pipes and valves. The diaphragm is used to circulate the gas between the atomizer and radon detector, the MFC is used to control the gas flow rate, and the vacuum pump is used to achieve a vacuum on the radon detector before gas transfer. 

To get a lower background, the atomizer, the water tank of the dehumidifier, and the radon detector are electro-polished to a roughness of 0.1~$\mu$m. Knife-edge flanges with metal gaskets, Vacuum Coupling Radius (VCR) connectors, and indium wire seal are used. The background event rate of the system is 3.0 $\pm$ 1.7 Counts Per Day (CPD), which corresponds to a one-day measurement of 1.0~mBq/m$^3$. The sensitivity at 90\% confidence level is estimated according to Eq.~\ref{Eq.L_c}~\cite{SK_2017}:
\begin{equation}
	L_c=\frac{1.64 \times \sigma_{BG}}{24 \times C_F}
	\label{Eq.L_c}
\end{equation}
where L$_c$ is the sensitivity in the unit of Bq/m$^3$, $\sigma_{BG}$ is the systemic uncertainty of the background in the unit of CPD, C$_F$ is the calibration factor in the unit of Counts Per Hour/(Bq/m$^3$) (CPH/(Bq/m$^3$)). The details of C$_F$ will be discussed in Sec.4.1.

\subsection{System operation}

\begin{figure}
	\centering
	\includegraphics[width=7cm]{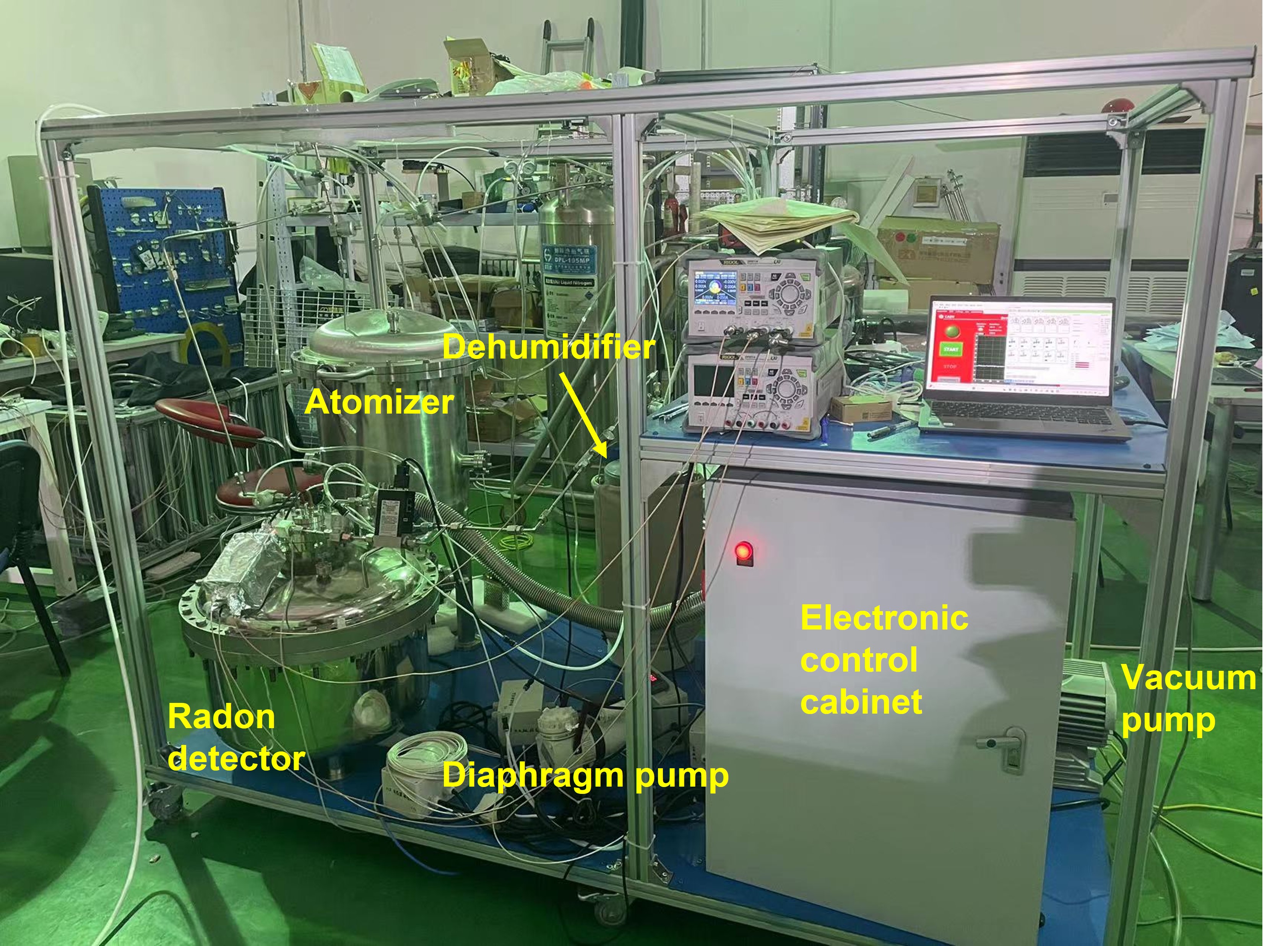}
	\caption{A picture of the radon concentration in water measurement system.}
	\label{fig.PLC}
\end{figure}

For remote operation, magnetic VCR valves for gas and magnetic stainless steel valves for water are used. Fig.~\ref{fig.PLC} shows a picture of the system. In addition, a Programmable Logic Controller (PLC) system was developed to remotely control the system components, including the MFC, the diaphragm pump, the vacuum pump, the dehumidifier, and the magnetic valves. 

\begin{figure}[htb]
	\centering
	\includegraphics[height=6cm]{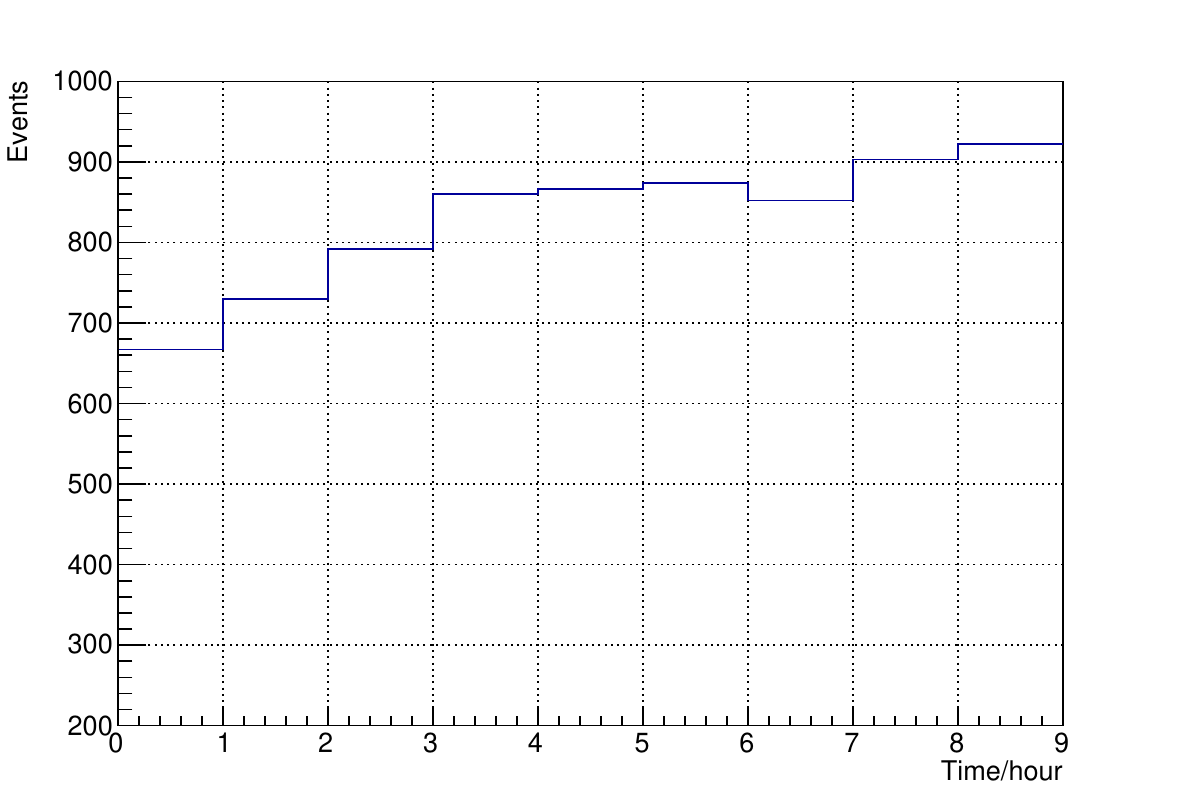}
	\caption{ Change of the $^{214}$Po event rate measured by the radon detector over time.}
	\label{water_equilibrium}
\end{figure}

When the radon concentration in water measurement system is operating in circulation mode, the diaphragm pump and dehumidifier are in operation, the MFC is set to 1~L/min, the relevant valves on the gas line are opened, and the radon detector is in data-taking mode. Fig.~\ref{water_equilibrium} shows the change in the measured counting rate of the radon detector over time. It is evident from the figure that in $\sim$5 hours the radon concentration in water and vapor as well as the radon concentration in the whole system could reach equilibrium. Based on Fig.~\ref{water_equilibrium}, we also can conclude that when this system is operated in gas transfer mode 5 hours of continuous work of the atomizer is sufficient before gas transfer.

\section{Results}
\subsection{Detector calibration }
The radon detector operates on the principle of electrostatic collection, and its detection efficiency must be calibrated before use. For this purpose, a gaseous $^{222}$Rn source, made from BaRa(CO$_3$)$_2$ powder by the radon laboratory of South China University~\cite{zhang_development_2018,chenYY_2022,Li_2023,liu_system_2023}, was employed for detector calibration. The Calibration Factor (C$_F$), as defined in Eq.\ref{Eq.CF}, is utilized to convert the measured event rate to the radon concentration in the gas. The radon concentration in the outgassing of the radon source is positively correlated with the gas flow rate. While calibration, the gas flow rate passing the radon source is set to 1~L/min and the radon concentration in the gas is 110.0 $\pm$ 7.2~Bq/m$^3$, which is measured by a calibrated RAD7 radon detector (Durridge Company).
\begin{equation}
	C_F = \frac{R_{Po-214}}{C_{Rn}}
	\label{Eq.CF}
\end{equation}
where R$_{Po-214}$ is the event rate of the measured $^{214}$Po event rate in the unit of CPH, C$_{Rn}$ is the radon concentration in the gas in the unit of Bq/m$^3$.

The detection efficiency can be derived from the C$_F$ using Eq.\ref{Eq.efficiency}. 
\begin{equation}
	\varepsilon =\frac{C_{Rn} \times V_D \times 3600~s \times 0.5}{C_F}
	\label{Eq.efficiency}
\end{equation}
where $\varepsilon$ is the detection efficiency, V$_D$ is the detector volume which is 0.1~m$^3$,  0.5 is because only the $\alpha$s emitted to the surface of the Si-PIN detector can be detected. 

\begin{figure}[htb]
	\centering
	\includegraphics[width=13.5cm]{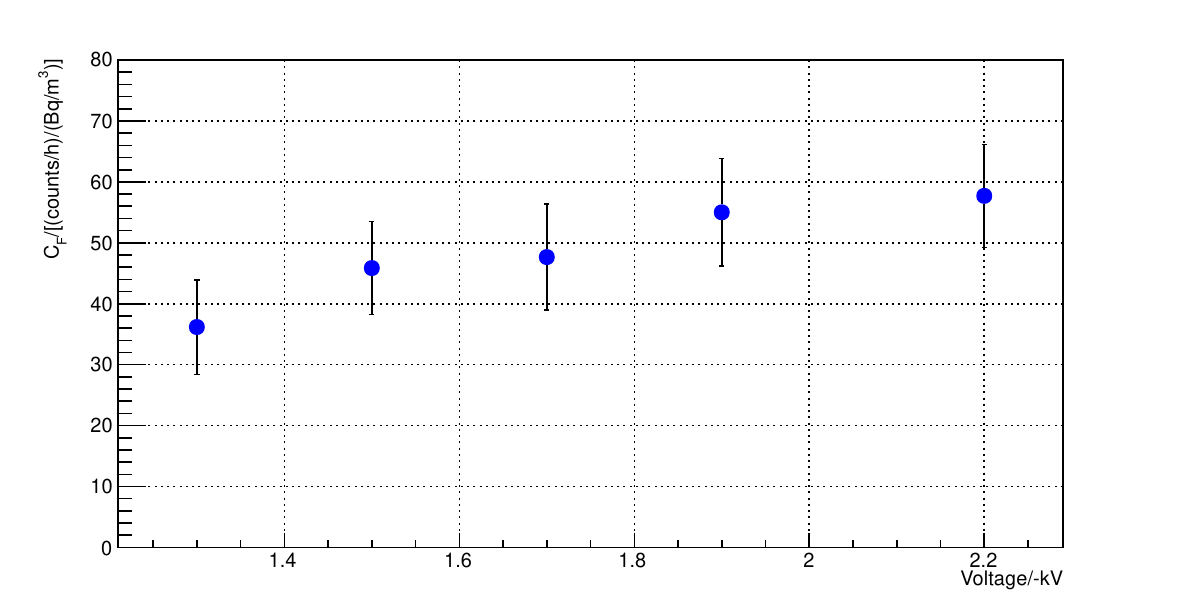}
	\caption{The high voltage dependence of the C$_F$. These data were obtained at an absolute humidity of 0.45~g/m$^{3}$. The errors include both statistical and systematic errors.}
	\label{High_voltage}
\end{figure}

For a specific detector, the detection efficiency is affected by the electric field within the detector and the humidity of the gas~\cite{SK_2020,Kotrappa1981ElectretaNT}. Fig.~\ref{High_voltage} illustrates the variation of C$_F$ with the high voltage applied to the Si-PIN detector. It is observed that C$_F$ increases with the rise in voltage. For this measurement, the absolute humidity was maintained at 0.45~g/m$^{3}$.

As depicted in Fig.~\ref{fig:radonsystem}, the radon source is positioned between the atomizer and the dehumidifier. Water remains present inside the atomizer, so the relative humidity of the gas coming out of the atomizer exceeds 50\%. The dehumidifier utilizes low temperatures for dehumidification, thus the humidity of the outgoing gas is relevant to the temperature of the dehumidifier. Lower temperatures yield improved dehumidification. Tab.~\ref{humidity} presents the relationship between the dehumidifier temperature, the humidity measured by the dew point meter, the calibration factor, and the detection efficiency. The detection efficiency of the detector rises as the gas humidity gradually decreases, with the collection efficiency being $\sim$68.5\% when the absolute humidity is 0.007~g/m$^3$.

\begin{table}[htb]
	\centering
	\renewcommand{\arraystretch}{1.5}
	\scriptsize
	\caption{ Relationship between detection efficiency of radon detector and gas humidity. During this measurement, the high voltage is set to -2200V. }
	\begin{tabular}{ccccc}
	\hline
	Dehumidifier Temperature  & Absolute humidity  & C$_F$  & Detection efficiency \\ 
    ($^{\circ}$C)  & (g/m$^{3}$) & (CPH/Bq/m$^{3}$) & (\%) \\
	\hline
	-40 &  0.45 & 62.0~±~4.1 &  34.4~±~2.3   \\
	-60 & 0.23 & 95.1~±~6.2  &  52.8~±~3.4   \\
	-80 & 0.04 & 100.0~±~6.5 &  55.6~±~3.6   \\
	-120 & 0.007 & 123.3~±~8.1 &  68.5~±~4.5  \\
	\hline 
	\label{humidity}	
	\end{tabular}
\end{table}

\subsection{Radon removal efficiency}

The radon concentration in water samples can be measured under various conditions by manipulating the bypass valves of the water system. In the subsequent experiments, we measured the radon concentrations in raw water, in degassed water with different stages of degasser, and in degassed water with three stages of degasser and one or two micro-bubble generators activated.  During the measurements, the transfer mode of the radon concentration measurement system is used. For the detector, the high voltage is set to -2200~V. For the dehumidifier, the temperature is set to -60~$^\circ$C. For the MFC, the gas flow rate is set to 1~L/min. Using this setup, the C$_F$ is 95.1 $\pm$ 6.2 ~CPH/(Bq/m$^3$). Before gas transfer, the atomizer was operated continuously for at least 5 hours.

\subsubsection{Radon concentration in raw water}
\begin{figure}[htb]
	\centering
	\includegraphics[width=11cm]{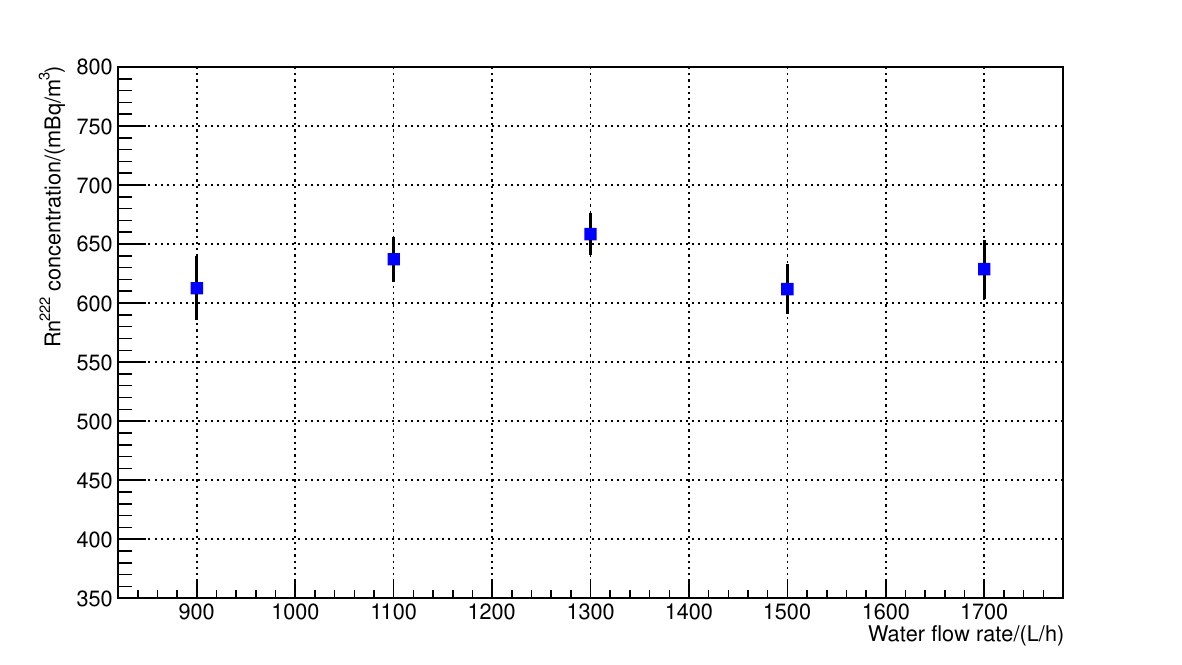}
	\caption{Measurement results of radon concentration in raw water at different water flow rates.}
	\label{Inlet_water_velocity}
\end{figure}

Fig.~\ref{fig:watersystem} shows the flow chart of the water system of the JUNO prototype. This water system has a maximum water flow rate of 1700~L/h, but the adjustment valve can control the water flow rate change between 900~L/h and 1700~L/h. The volume of the JUNO prototype water tank is 5~m$^3$, therefore it takes 3-5 hours to circulate one volume.  To determine the radon concentration in water and to verify that devices other than the degassers and microbubble generators in the water system do not affect the radon concentration in water, raw water with different water flow rates is measured, and the results are shown in Fig.~\ref{Inlet_water_velocity}.

According to the results presented in Fig.~\ref{Inlet_water_velocity}, it is evident that the radon concentration in the water remains stable at $\sim$620~mBq/m$^3$ when the degassing membrane and microbubble devices are inactive. 

\subsection{System optimization}

The radon concentration in the water decreases after degassing, and the degassed water is returned directly to the tank during circulation. Consequently, the radon concentration in water decreases slightly as it circulates during one experiment. As a result, accurate estimation of the radon removal efficiency of the degasser becomes challenging. To address this, a gas flow radon source with evaporated nitrogen as the gas source is used to continuously introduce radon-containing gas into the tank, thereby ensuring the stability of the radon concentration in the water. By adjusting the gas flow rate, the radon concentration in the water was varied within the range of 300~mBq/m$^3$ to 10~Bq/m$^3$.

In the process of radon removal from ultra-pure water, it is observed that with an increasing number of cycles, the gas content in the water decreases, resulting in a decrease in the efficiency of radon removal by the degassing membrane and the presence of a removal limit. To investigate this phenomenon, researchers conduct multiple cycles of degassing on ultra-pure water with an initial radon concentration of $\sim$1~Bq/m$^{3}$. The experimental results showed that the radon concentration in the ultra-pure water reached a plateau at $\sim$0.25 Bq/m$^{3}$, indicating the attainment of the removal limit. Consistent results were reported in our previous work~\cite{guo_study_2018}. This is mainly because, after many rounds of degassing, the gas concentration in the water is already extremely low, so it cannot be degassed further. To increase the gas concentration in water, the microbubble generator is used. As is shown in Fig.~\ref{fig:watersystem}, two microbubble generators are used, Micro-bubble-1 is in the primary water line and Micro-bubble-2 is in the bypass. Micro-bubble-2 could inject gas-loaded water into the system after the first stage of degasser. 

\begin{figure}[htb]
	\centering
	\includegraphics[width=11cm,height=6.5cm]{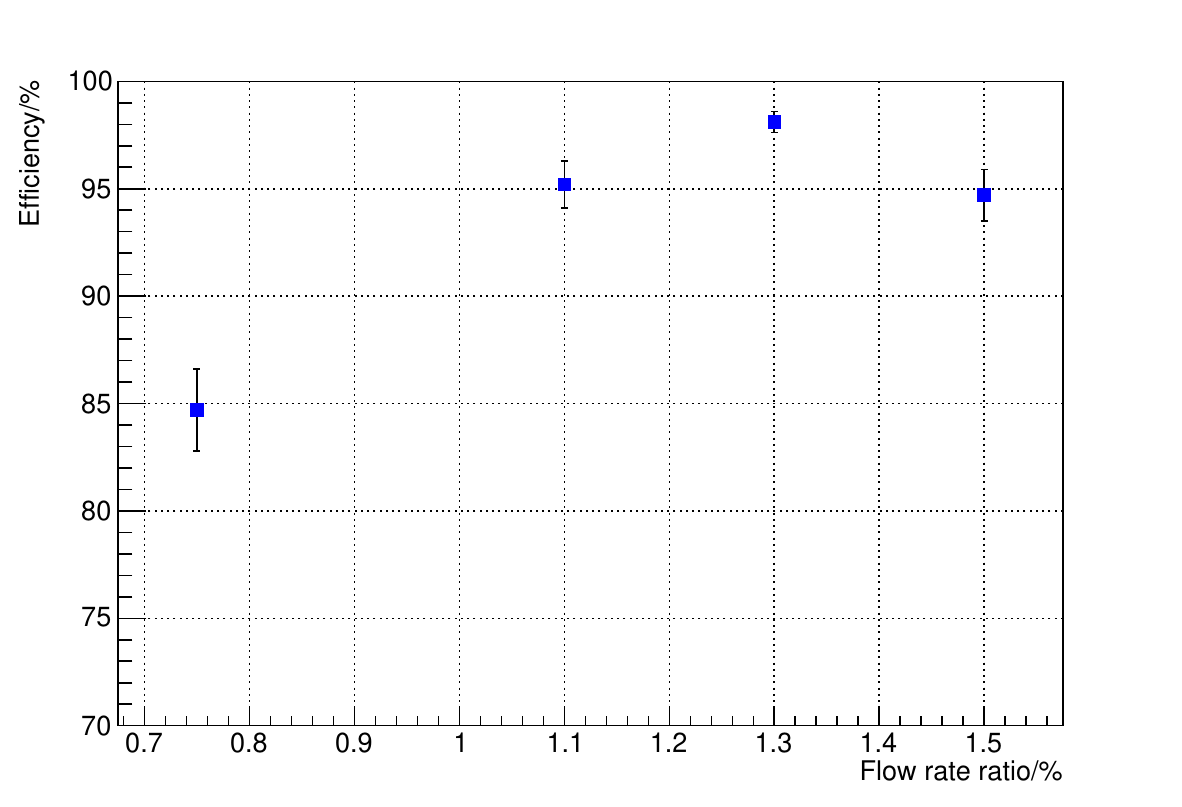}
	\caption{\label{flow_rate}Relationship between radon removal efficiency and inlet flow rate of microbubble generator.}
\end{figure}

In the conventional use of the microbubble device, the water outlet of the micro-bubble generator needs to be connected to a container without water pressure, while in this experiment, the outlets of the micro-bubble generators need to be connected to the water system pipeline, and the water pressure in the pipeline is about 0.2 MPa, so we need to optimize the microbubble device, that is, to adjust the ratio of the gas flow rate to the inlet water flow rate. The results are shown in Fig.~\ref{flow_rate}. It can be seen that the system is most efficient when the inlet gas flow is $\sim$1.5\% of the inlet water flow.

Subsequently, the radon concentrations in water at various conditions have been measured. The results are summarized in Tab.~\ref{results}. The degassing efficiency of the three stages of degassers is slightly better than that of one single stage, while the introduction of a micro-bubble generator can greatly improve the degassing efficiency. For ultrapure water with an initial radon concentration of 969 $\pm$ 63~mBq/m$^3$, the radon removal efficiency of the system can reach $\sim$99\% when two microbubble devices are used for gas loading and three-stage degassing membranes are used for degassing.

\begin{table}[htb]
	\centering
	\renewcommand{\arraystretch}{1.5}
	\scriptsize
	\caption{\label{results} Measurement results of radon concentration in water at different conditions.}
	\begin{tabular}{ccccccc}
	\hline
        Degasser-1 & Degasser-2 & Degasser-3 & Microbubble-1 & Microbubble-2 &  Efficiency (\%)\\
        ON & OFF & OFF & OFF & OFF & 50.9 $\pm$ 4.8  \\
        ON & ON & ON & OFF & OFF & 68.1 $\pm$ 3.3 \\
        ON & ON & ON & ON  & OFF &  94.99 $\pm$ 0.62 \\
        ON & ON & ON & ON & ON &  98.83 $\pm$ 0.27\\ \hline
	\end{tabular}
\end{table}

\subsubsection{Radon removal efficiency for various radon concentration water}

Under the optimal conditions mentioned above, we investigated the radon removal efficiency of the system at various radon concentrations, and the results are shown in Tab.~\ref{Efficiency}. For the radon concentration in water between $\sim$340~mBq/m$^3$ to $\sim$10~Bq/m$^3$, the radon removal efficiency is $\sim$98.5\% for a single pass.  

\begin{table}[htb]
	\centering
	\renewcommand{\arraystretch}{1.5}
	\scriptsize
	\caption{Radon removal efficiency measurement results of water circulation purification system under optimal conditions.}
	\begin{tabular}{ccccc}
		\hline
		& Raw water(mBq/m$^{3}$) & Degassed water(mBq/m$^{3}$) & Efficiency(\%) \\
		\hline 
		1 & 9126~±~595 & 157~±~24 & 98.28~±~0.29 \\
		2 & 4578~±~298 & 76.0~±~8.6 & 98.34~±~0.22  \\
		3 & 1329~±~91 & 25.2~±~4.8 & 98.10~±~0.38  \\
		4 & 969~±~75 & 11.3~±~2.5 & 98.83~±~0.27 \\
		5 & 665~±~53 & 8.1~±~2.3 & 98.78~±~0.35  \\
		6 & 560~±~45 & 10.5~±~2.6 & 98.12~±~0.49  \\
        7 & 338 $\pm$ 25 & 3.6 $\pm$ 2.3 & 98.93 $\pm$ 0.69\\
		\hline 	
    \label{Efficiency}
	\end{tabular}
\end{table}

\subsubsection{Radon removal limit}
For the JUNO experiment, the water system will be utilized for online radon removal, necessitating the circulation of water from the WCD through the water system multiple times. In this experiment, we circulated the water, with an initial radon concentration of 637 $\pm$ 51~mBq/m$^3$, under the optimal conditions for 25 hours at a water flow rate of 1.3~m$^3$/h, equivalent to $\sim$6.5 purification cycles. Following one-day measurement, a radon concentration in water of 0.66 $\pm$ 0.66~mBq/m$^3$ is obtained, the errors include both systematic and statistical errors. Despite the substantial error associated with this result, it is essentially certain that the system can produce ultrapure water with a radon concentration of sub-mBq/m$^3$.

\section{Summary}
JUNO is a multi-purpose neutrino experiment and its main physics goal is to determine the neutrino mass ordering. According to the Monte Carlo simulation results, the radon concentration in water has to be reduced to less than 10~mBq/m$^3$. To support the research and development of low-radon ultrapure water, an ultrapure water circulation system and a high-sensitivity online radon concentration measurement system, capable of detecting concentrations as low as 1~mBq/m$^3$, have been established based on the JUNO prototype. The water system incorporates Liquid-Cel degassing membranes for radon gas removal, and microbubble generators are used to introduce gas into the water to enhance radon removal efficiency. By employing two microbubble generators for gas-loading and implementing three stages of degassers for gas extraction, the radon removal efficiency can reach approximately 98.5\%. In a single pass, this system can reduce the radon concentration from $\sim$1~Bq/m$^3$ to $\sim$10~mBq/m$^3$, and with multiple cycles, the system can produce ultrapure water with radon concentrations at a sub-mBq/m$^3$ level. This system could meet the stringent requirements of JUNO.
	
\section{Acknowledgement}
This work is supported by the State Key Laboratory of Particle Detection and Electronics (Grant No. SKLPDE-ZZ-202304), the Youth Innovation Promotion Association of the Chinese Academy of Sciences (Grant No. 2023015), and the Strategic Priority Research Program of the Chinese Academy of Sciences (GrantNo.XDA10011200).


\begin{thebibliography}{}
\bibitem{adam_juno_2015}A. Abusleme et al., JUNO Collaboration, JUNO physics and detector, Prog. Part. Nucl. Phys. 123 (2022) 103927. http://dx.doi.org/10.1016/j.ppnp.2021.103927.
\bibitem{noauthor_juno_2022}F. An et al., JUNO Collaboration, Neutrino physics with JUNO, J. Phys. G: Nucl. Part. Phys. 43 (2016) 030401, 10.1088/0954-3899/43/3/030401.
\bibitem{dayabay} F.P. An et al., Daya Bay Collaboration, Observation of electron-antineutrino disappearance at Daya Bay, Phys. Rev. Lett. 108 (2012) 171803, DOI:10.1103/PhysRevLett.108.171803.
\bibitem{resin} https://www.lenntech.com/Data-sheets/Dowex-Monosphere-MR-3-UPW-L.pdf.
\bibitem{Li-cel} https://www.3m.com/3M/en$\_$US/liquicel-us/.
\bibitem{membrane_2017}Joel Minier-Matar et al. Application of membrane contactors to remove hydrogen sulfide from sour water.Journal of Membrane Science 541 (2017) 378-385. https://doi.org/10.1016/j.memsci.2017.07.025.
\bibitem{guo_study_2018}C. Guo et al. Study on the radon removal for the water system of Jiangmen Underground Neutrino Observatory. RDTM 2 (2018) 48, https://doi.org/10.1007/s41605-018-0077-8. 
\bibitem{MB-1} W. Zimmerman, V. Tesar, S. Butler, H. Bandulasena, Microbubble generation, Recent. Pat. Eng. 2 (2008) 1-8.
\bibitem{MB-2} J. Rodríguez-Rodríguez, A. Sevilla, C. Martínez-Bazán, J.M. Gordillo, Generation of microbubbles with applications to industry and medicine, Annu. Rev. Fluid Mech. 47 (2015) 405-429.
\bibitem{2012Microbubble}Snigdha Khuntia, Subrata Kumar Majumder, and Pallab Ghosh. Microbubble-aided water and wastewater purification: a review. Reviews in Chemical Engineering, 28 (2012) 191-221. https://doi.org/10.1515/revce-2012-0007.
\bibitem{2006Ozone}Pan Li and Hideki Tsuge. Ozone Transfer in a New Gas-Induced Contactor with Microbubbles. JOURNAL OF CHEMICAL ENGINEERING OF JAPAN, 39 (2006) 1213-1220. https://doi.org/10.1252/jcej.39.1213.
\bibitem{liu_degradation_2018} Yanan Liu, et al. Degradation of aniline in aqueous solution using non-thermal plasma generated in microbubbles. Chemical Engineering Journal 345 (2018) 679-687. https://doi.org/10.1016/j.cej.2018.01.057	
\bibitem{khan_micronanobubble_2020}Khan P et al., Micro-nanobubble technology and water-related application. Water Supply 20 (2020) 2021-2035. https://doi.org/10.2166/ws.2020.121.
\bibitem{liu_system_2023}Y. Liu et al. System upgrade for $\mu$Bq/m$^{3}$ level $^{222}$Rn concentration measurement, JINST 18 (2023) T03002, https://doi.org/10.1088/1748-0221/18/03/T03002.
\bibitem{Weigel} Weigel, F., Radon. Chem. Ztg. 102 (1978) 287-299.
\bibitem{RM} Lucchetti Carlo et al., Testing the radon-in-water probe set-up for the measurement of radon in water bodies, Radiation Measurements 128 (2019) 106179, https://doi.org/10.1016/j.radmeas.2019.106179.
\bibitem{JER} Thomas C. Stieglitz, Peter G. Cook, William C. Burnett, Inferring coastal processes from regional-scale mapping of 222Radon and salinity: examples from the Great Barrier Reef, Australia, Journal of Environmental Radioactivity 101 (2010) 544–552, doi:10.1016/j.jenvrad.2009.11.012.
\bibitem{SuperK1999} Y. Takeuchi et al., Development of high sensitivity radon detectors, NIM A 421 (1999) 334-341.
\bibitem{10.1093/ptep/ptv018}K. Hosokawa et al., Development of a high-sensitivity 80 L radon detector for purified gases. Progress of Theoretical and Experimental Physics 2015 (2015) 033H01. https://doi.org/10.1093/ptep/ptv018.
\bibitem{SNO} Jian-Xiong Wang, Tom C. Andersen, John J. Simpson, An electrostatic radon detector designed for water
radioactivity measurements. NIM A 421 (1999) 601-609.
\bibitem{zhang_development_2018}Y.P Zhang, et al. The development of $^{222}$ Rn detectors for {JUNO} prototype. RDTM 2 (2018) 5. https://doi.org/10.1007/s41605-017-0029-8.
\bibitem{chenYY_2022}Y.Y Chen et al. A study on the radon removal performance of low background activated carbon, JINST 17 (2022) P02003, https://doi.org/10.1088/1748-0221/17/02/P02003.
\bibitem{Li_2023}C, Li, et al. Study on the radon adsorption capability of low-background activated carbon. J Radioanal Nucl Chem (2023). https://doi.org/10.1007/s10967-023-09211-w.
\bibitem{HV_tianjin} http://tjcentre.com.cn/.
\bibitem{wang_SiPM_2021}L.F. Xie et al., Developing the radium measurement system for the water Cherenkov detector of the Jiangmen Underground Neutrino Observatory, NIM A 976 (2020) 164266.  Characterization of VUV4 SiPM for liquid argon detector.JINST 16 (2021) P07021, http://doi.org/10.1088/1748-0221/16/07/P07021, https://doi.org/10.1016/j.nima.2020.164266.
\bibitem{SK_2017}Y. Nakano et al. Measurement of radon concentration in super-Kamiokande’s buffer gas. NIM A 867 (2017) 108-114, http://dx.doi.org/10.1016/j.nima.2017.04.037.
\bibitem{SK_2020}Y. Nakano et al. Measurement of the radon concentration in purified water in the {Super}-{Kamiokande} {IV} detector. NIM A 977 (2020) 164297, https://doi.org/10.1016/j.nima.2020.164297.
\bibitem{Kotrappa1981ElectretaNT}Kotrappa, Paul et al. Electret--a new tool for measuring concentrations of radon and thoron in air. Health Physics 41 (1981) 35-46, https://doi.org/10.1097/00004032-198107000-00004.

\end{thebibliography}
\end{document}